\documentclass[
10pt,
showpacs,preprintnumbers,nofootinbib,
 amsmath,amssymb,
 aps,
prl,twocolumn,groupedaddress,superscriptaddress,
showkeys
]{revtex4-1}
\usepackage{graphicx}
\usepackage{dcolumn}
\usepackage{bm}
\usepackage[colorlinks=true,urlcolor=blue,citecolor=blue,breaklinks=true]{hyperref}
\usepackage{epsfig}

\begin{document}


\title{$^4{\rm He}$+$n$+$n$ continuum within an {\em ab initio} framework}

\author{Carolina Romero-Redondo}
 \email{cromeroredondo@triumf.ca}
\affiliation{TRIUMF, 4004 Wesbrook Mall, Vancouver, British Columbia, V6T 2A3, Canada}
\author{Sofia Quaglioni}
 \email{quaglioni1@llnl.gov}
 \affiliation{Lawrence Livermore National Laboratory, P.O. Box 808, L-414, Livermore, California 94551, USA}
 \author{Petr Navr\'atil}
 \email{navratil@triumf.ca}
 \affiliation{TRIUMF, 4004 Wesbrook Mall, Vancouver, British Columbia, V6T 2A3, Canada}
\author{Guillaume Hupin}
 \email{hupin1@llnl.gov}
 \affiliation{Lawrence Livermore National Laboratory, P.O. Box 808, L-414, Livermore, California 94551, USA}

\date{\today}

\begin{abstract}
The low-lying continuum spectrum of the $^6{\rm He}$ nucleus is investigated for the first time within an {\em ab initio} framework  that encompasses the  $^4{\rm He}$+$n$+$n$  three-cluster dynamics characterizing its lowest decay channel. This is achieved through an extension of the no-core-shell model combined with the resonating-group method, in which energy-independent non-local interactions among three nuclear fragments can be calculated microscopically starting from realistic nucleon-nucleon interactions and consistent {\em ab initio} many-body wave functions of the clusters.   The three-cluster Schr\"odinger equation is solved with three-body scattering boundary conditions by means of the hyperspherical-harmonic method on a Lagrange mesh.  Using a soft similarity-renormalization-group evolved chiral nucleon-nucleon potential, we find the known $J^\pi = 2^+$ resonance as well as a result consistent with a new low-lying second $2^+$ resonance recently observed at GANIL at $2.6$ MeV above the $^6$He ground state. 
We also find resonances in the $2^-$, $1^+$ and $0^-$ channels, while no low-lying resonances are present in the $0^+$ and $1^-$ channels.
\end{abstract}
\maketitle

%
\paragraph{Introduction.}
Nuclear systems near the drip lines, the limits of the nuclear chart beyond which neutrons or protons start dripping out of nuclei, offer an exciting opportunity to advance our current understanding of the interactions among nucleons, so far  mostly based on the study of stable nuclei. This is not a goal devoid of challenges.
Experimentally, 
the study of these rare nuclei with atypical neutron-to-proton ratios is challenged by their short half lives and minute production cross sections. A major stumbling block in nuclear theory has to deal with the low breakup thresholds, which cause bound, resonant and scattering states to be strongly coupled. Particularly arduous in this respect are those systems for which the lowest threshold for particle decay is of the three-body nature, such as $^6$He, which decays into an $\alpha$ particle ($^4$He nucleus) and two neutrons at the excitation energy of 0.975 MeV. Aside from a narrow resonance characterized by spin-parity $J^\pi=2^+$, located at 1.8 MeV above the ground state (g.s.), the positions, spins and parities of the excited states of this nucleus are still under discussion. Experimentally the picture is not clear. Proton-neutron exchange reactions between two fast colliding nuclei produced resonant-like structures around $4$~\cite{PhysRevLett.85.262} and 5.6~\cite{PhysRevC.54.1070} MeV of widths $\Gamma\sim4$ and $10.9$ MeV, respectively, as well as a broad asymmetric bump at $\sim5$ MeV~\cite{Nakamura2000209}, but disagree on the nature of the underlying $^6$He excited state(s). While the structures of Refs.~\cite{PhysRevLett.85.262} and~\cite{Nakamura2000209} are explained as  dipole excitations  compatible 
 with oscillations of the positively-charged $^4$He core against the halo neutrons,
 that of Ref.~\cite{PhysRevC.54.1070} is identified as a second $2^+$ state. More recently, a much narrower $2^+$ ($\Gamma=1.6$ MeV) state and a $J=1$ resonance ($\Gamma\sim2$ MeV) of unassigned parity  were populated at $2.6$ and $5.3$ MeV, respectively, with the two-neutron transfer reaction $^8$He($p,^3$H)$^6$He$^*$~\cite{Mougeot:2012aq}. On the theory side, several predictions all incomplete
in different ways suggest a $2_1^+,2_2^+,1^+,0^+$ sequence of levels above the first excited state but disagree on the positions and widths. Those from six-body calculations with realistic Hamiltonians~\cite{PRL.87.172502,PhysRevC.68.034305,PhysRevC.70.054325} were obtained within a bound-state approximation and cannot provide any information about the widths of the levels. {\em Vice versa}, those from three-body models~\cite{PhysRevC.55.R577,Danilin1998383}, from  microscopic three-cluster models~\cite{Korennov2004249, PhysRevC.80.044310} 
or from calculations hinging on a shell-model picture with inert $^4$He core~\cite{PhysRevC.71.044314, PhysRevC.76.054309} can describe the continuum, but were obtained using schematic interactions and a simplified description of the structure.
In this Letter we present the first {\em ab initio} calculation of the $^4{\rm He}$+$n$+$n$ continuum starting from a nucleon-nucleon ($NN$) interaction that describes two-nucleon properties with high accuracy. 

\paragraph{Formalism.}
In the no-core shell model combined with the resonating-group method (NCSM/RGM), 
$A$-body bound and/or scattering states characterized by three-cluster configurations are described by the wave function 
\begin{equation}
|\Psi^{J^{\pi}T}\rangle=\sum_{\nu}
\!\!\iint \!dx \, dy\, x^2y^2 \, 
\hat{\mathcal A}_{\nu}
|\Phi_{\nu xy}^{J^{\pi}T}\rangle\, 
\big [{\mathcal N}^{-\frac12}\chi\big]^{J^{\pi}T}_\nu(x,y), 
\label{eq:ABodyWF}
\end{equation}
in terms of $(A-a_{23},a_2,a_3)$ ternary cluster channels 
\begin{align}
&|\Phi_{\nu xy}^{J^{\pi}T}\rangle \label{eq:basis} \\
&\!=\!\Big[\!\big (|A-a_{23} \alpha_1 I^{\pi_1}_1 T_1\rangle
\!(|a_2 \alpha_2 I^{\pi_2}_2 T_2\rangle 
\!|a_3 \alpha_3 I^{\pi_3}_3 T_3\rangle 
)^{(s_{23}T_{23})}\!\big)^{(ST)}\nonumber\\
&\times( Y_{\ell_x}(\hat{\eta}_{23})Y_{\ell_y}(\hat{\eta}_{1,23}))^{(L)}\!\Big ]^{(J^{\pi}T)}
\frac{\delta(x-\eta_{23})}{x\eta_{23}}
\frac{\delta(y-\eta_{1,23})}{y\eta_{1,23}}\nonumber
\end{align}
built within a translation-invariant harmonic oscillator (HO) basis from NCSM eigenstates of each of the three clusters, $|A-a_{23} \alpha_1 I^{\pi_1}_1 T_1\rangle$, 
$|a_2 \alpha_2 I^{\pi_2}_2 T_2\rangle$ and 
$|a_3 \alpha_3 I^{\pi_3}_3 T_3\rangle$, and antisymmetrized with an appropriate operator $\hat{\mathcal A}_\nu$ to preserve the Pauli exclusion principle exactly. 
 Here $A-a_{23}, a_2$, and $a_3$ (with $A\ge a_{23}=a_2+a_3$) indicate the mass numbers of the three clusters having angular momentum, parity, isospin and energy quantum numbers $I_{i}^{\pi_i}T_i$ and $\alpha_i$ $(i = 1,2,3)$. 
Each channel is identified by its total isospin, angular momentum and parity 
($J^{\pi}T$) and an index $\nu$ specifying all other 
quantum numbers,
i.e., $\nu = \{A-a_{23}\,\alpha_1I_1^{\pi_1}T_1;$ $ a_2\, \alpha_2 I_2^{\pi_2} T_2; a_3\, \alpha_3 I_3^{\pi_3}T_3; s_{23} \,T_{23}\, S \,\ell_x \,\ell_y \, L\}$.
Further,
$\vec\eta_{1,23}$ $=\eta_{1,23}\hat\eta_{1,23}$ and $\vec\eta_{23}=\eta_{23}\hat\eta_{23}$ 
are relative coordinates 
proportional, respectively, to the displacement between the center of mass (c.m.)\ of the first cluster and that of the residual two fragments,
and to the distance between the 
c.m.'s of clusters 2 and 3.
 
Introducing the hyperspherical coordinates $\rho = \sqrt{x^2+y^2}$ and $\alpha = \arctan{\tfrac xy}$, 
the relative motion wave functions among the clusters, 
\begin{align}
	\chi_{\nu}^{J^{\pi}T}(\rho,\alpha) = \frac {1}{\rho^{5/2}}\sum_K u^{J^\pi T}_{\nu K}(\rho) \phi^{\ell_x,\ell_y}_K(\alpha)\,,
\label{hyper_expansion}
\end{align}
can be expanded over the complete set $\phi^{\ell_x,\ell_y}_K(\alpha)$, the hyperangular part of the hyperspherical harmonics 
$\mathcal{Y}^{K \ell_x\ell_y}_{L M_L}(\Omega)=\phi_K^{\ell_x,\ell_y}(\alpha) \left(
Y_{\ell_x}(\hat\eta_{23})\, Y_{\ell_y}(\hat\eta_{1,23})\right)^{(L)}_{M_L}$. 
The unknown amplitudes $u^{J^\pi T}_{\nu K}(\rho)$ 
are then found by solving the nonlocal hyperradial equations   \begin{align}
        \sum_{K\nu}\!\int \!\!d\rho \rho^5 {\cal \bar{ H}}_{\nu'\nu}^{K'K}(\rho',\rho) \frac{u^{J^{\pi}T}_{K\nu}(\rho)}{\rho^{5/2}} 
        = E \frac{u^{J^{\pi}T}_{K^\prime\nu^\prime}(\rho^\prime)}{\rho^{\prime\,5/2}}\,,
        \label{RGMrho}
\end{align}
where  ${\cal \bar{ H}}_{\nu'\nu}^{K'K}(\rho',\rho)= \big[{\cal N}^{-\frac12}{\cal H}\,{\cal N}^{-\frac12}\big]_{\nu'\nu}^{K'K}(\rho',\rho)$ 
is the orthogonalized kernel obtained from the Hamiltonian and overlap (or norm) matrix elements
\begin{align}
        {\mathcal H}^{J^\pi T}_{\nu^\prime\nu}(x^\prime,y^\prime,x,y) & = 
                \left\langle\Phi^{J^\pi T}_{\nu^\prime x^\prime y^\prime} \right| \hat {\mathcal A}_{\nu^\prime} H \hat {\mathcal A}_\nu \left | \Phi^{J^\pi T}_{\nu x y} \right\rangle\,, \label{eq:Hkernel} \\
        {\mathcal N}^{J^\pi T}_{\nu^\prime\nu}(x^\prime,y^\prime,x,y) & = 
                \left\langle\Phi^{J^\pi T}_{\nu^\prime x^\prime y^\prime} \right| \hat{\mathcal A}_{\nu^\prime}  \hat {\mathcal A}_\nu \left | \Phi^{J^\pi T}_{\nu x y} \right\rangle \,,  \label{eq:Nkernel}
\end{align}
after projection over the basis $\phi^{\ell_x,\ell_y}_K(\alpha)$.
Equation~(\ref{RGMrho}) is solved with either bound- or genuinely three-body scattering-state (i.e.\ no bound two-body subsystems are present) boundary conditions by means of the microscopic $R$-matrix method on a Lagrange
mesh \cite{PhysRevA.65.052710,BayeJPB98,Hesse2002184,Hesse199837,Descouvemont:2003ys}.
For more details on the three-cluster NCSM/RGM formalism we refer the interested reader to Ref.~\cite{PhysRevC.88.034320}, where  we first applied the approach to the description of the ground state of $^6$He within a $^4{\rm He}$+$n$+$n$ basis ($a_2,a_3=1$). Here we apply the same framework 
to the much more challenging problem of 
the continuum of this system.

\paragraph{Results.}
The present calculations are based on the chiral N$^3$LO $NN$~\cite{N3LO} interaction softened via the Similarity Renormalization Group (SRG)  to minimize the influence of momenta higher than $\Lambda$=1.5 fm$^{-1}$. This soft potential permits us to
reach convergence in the  HO expansions within $N_{\rm max}\sim 13$ quanta, the largest model space presently achievable.
At the same time, it also leads to a $^4$He g.s.\ energy~\cite{PhysRevLett.103.082501,PhysRevC.83.034301} and $n$+$^4$He phase shifts~\cite{PhysRevLett.108.042503} close to experiment despite the omission of three-nucleon ($3N$) forces, which are beyond the scope of this first {\em ab initio} study of the $^4$He+$n$+$n$ continuum.

\begin{figure}[t]
\includegraphics[width=75mm]{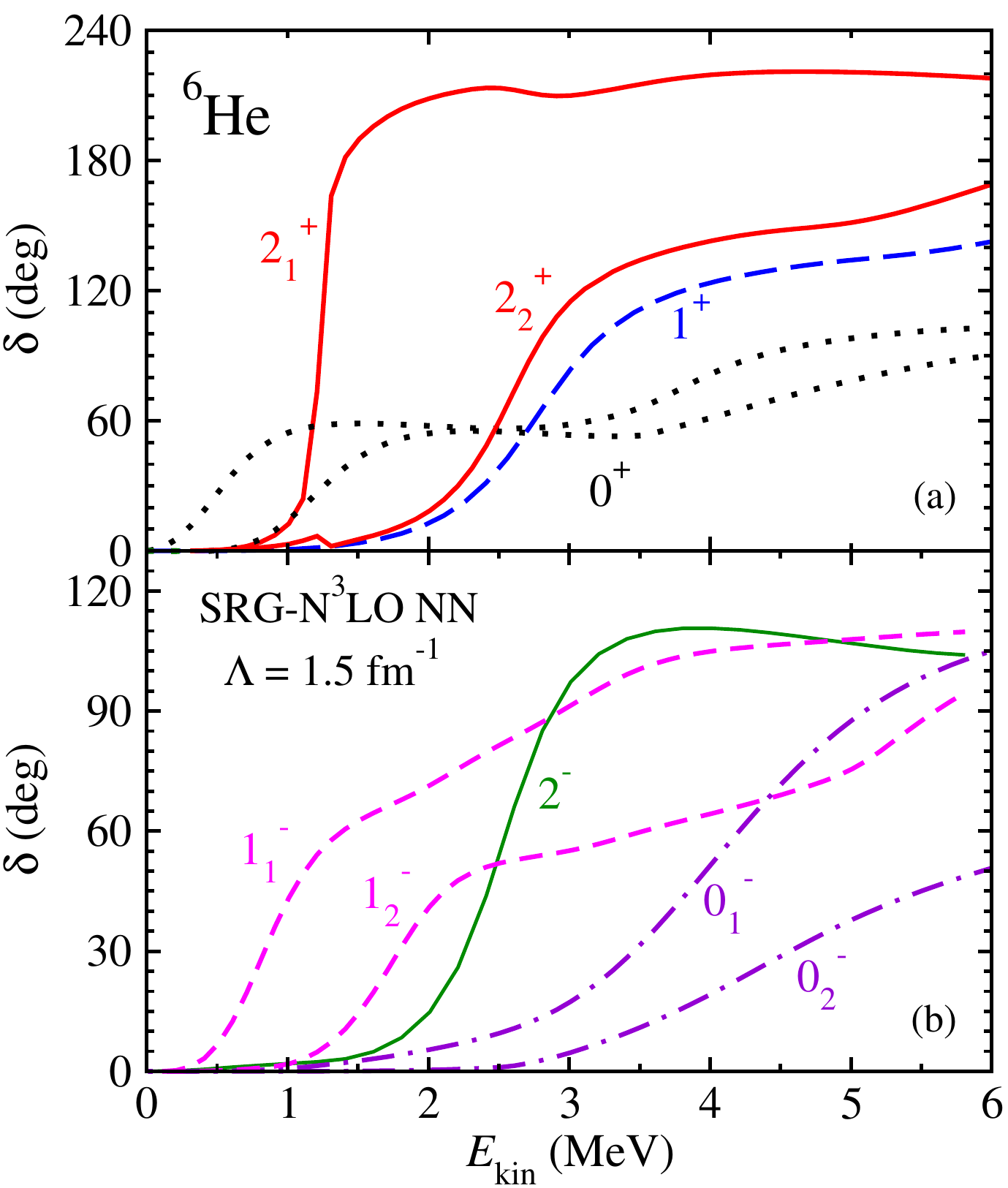}
\caption{(Color online) Calculated $^4{\rm He}$+$n$+$n$ (a) positive- and (b) negative-parity attractive eigenphase shifts (except for the $2^+$ and $1^+$ channels, where the diagonal phase shifts are shown) as a function of the three-cluster kinetic energy in the c.m.\ frame $E_{\rm kin}$.
See the text for further details.}
\label{fig:final}
\end{figure}
\begin{figure}[b]
\includegraphics[width=80mm]{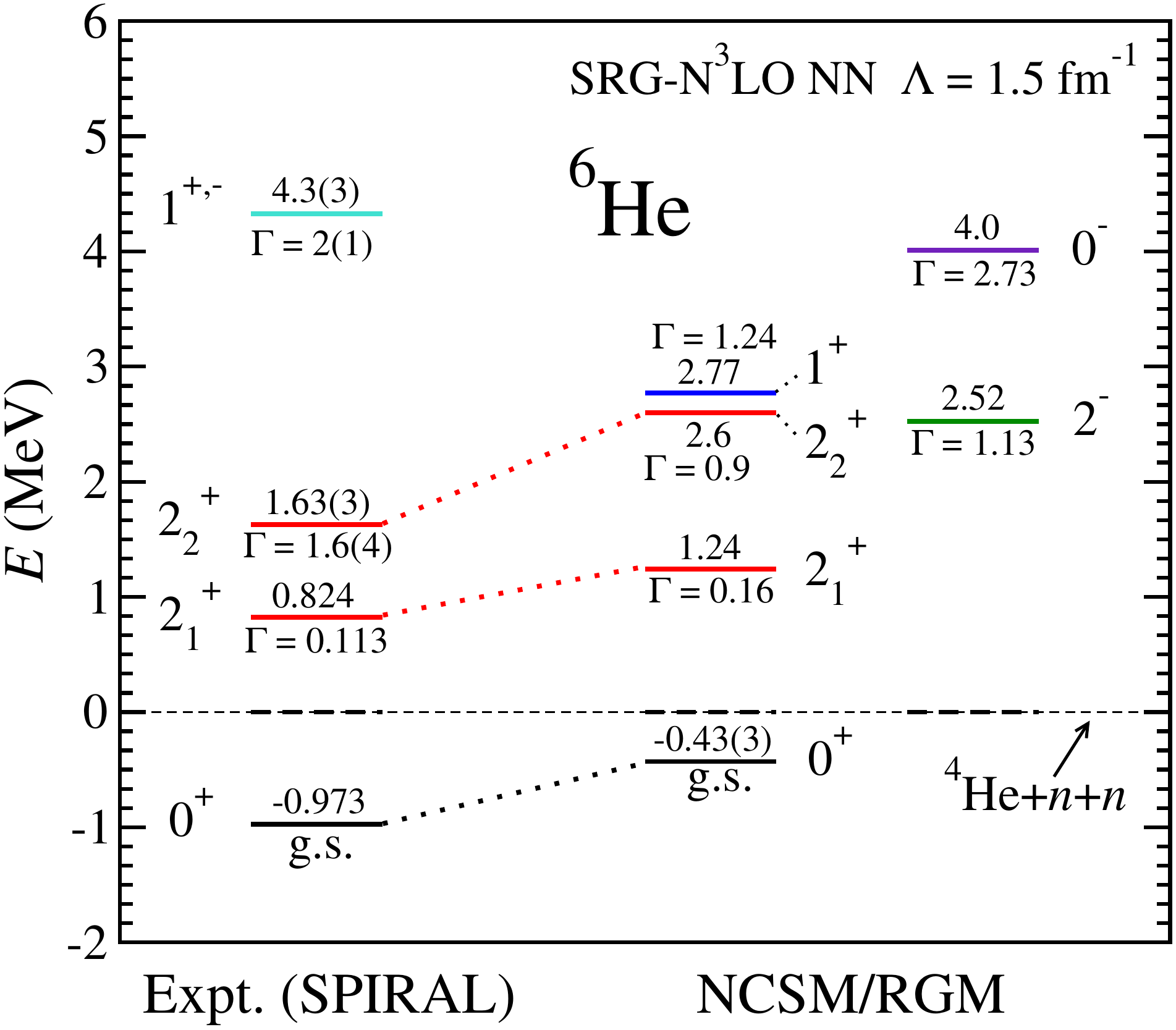}
\caption{(Color online). Comparison of the spectrum obtained within this work 
using the NCSM/RGM to the experimental spectrum measured at the SPIRAL facility (GANIL)~\cite{Mougeot:2012aq}.}
\label{fig:spec}
\end{figure}
\begin{figure*}[t]
    \begin{minipage}[c]{8.0cm}\hspace*{-6mm}
      \includegraphics[width=7.5cm,clip=,draft=false]{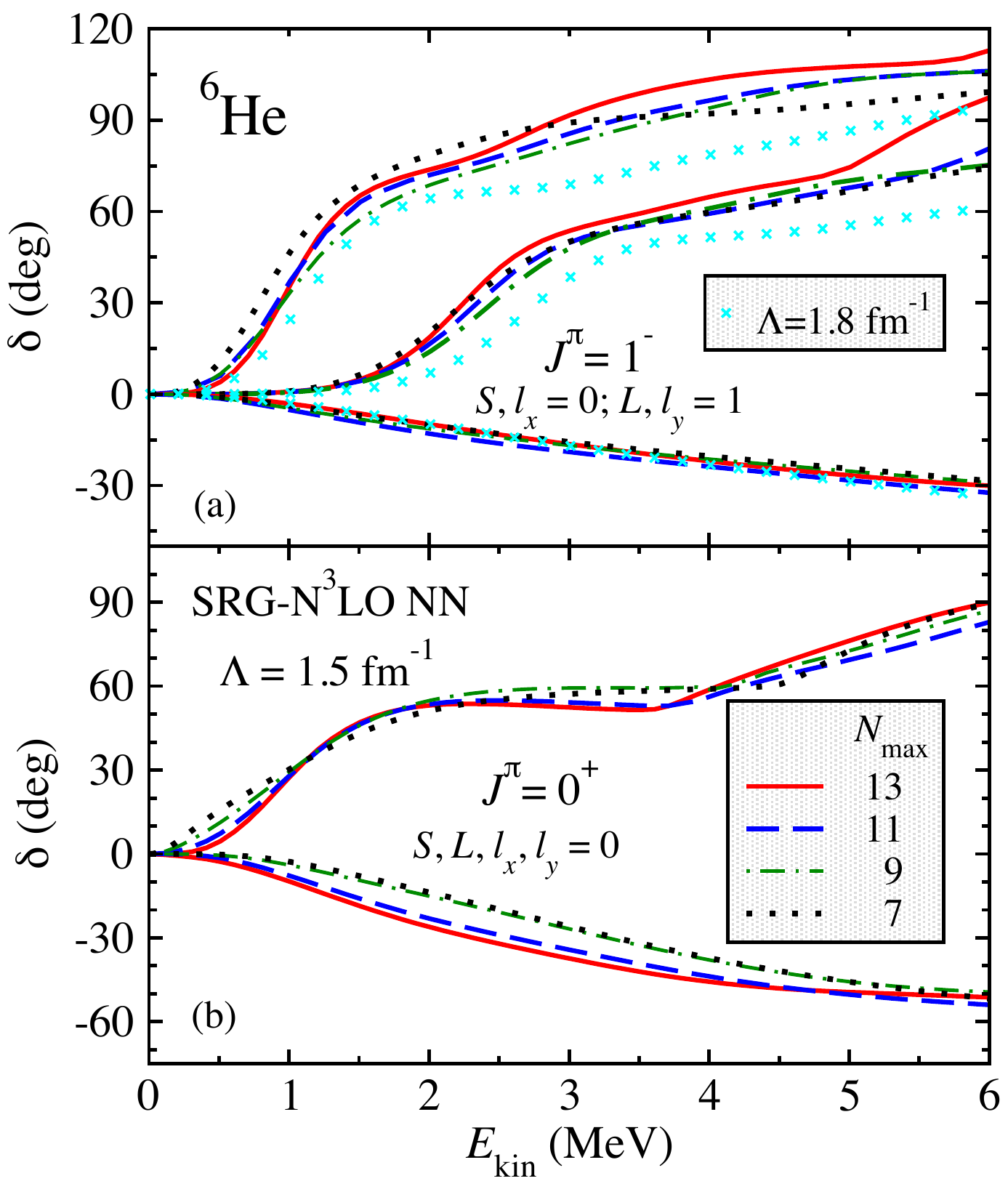}
    \end{minipage}
    \begin{minipage}[c]{8.0cm}\hspace*{-6mm}
      \includegraphics[width=7.5cm,clip=,draft=false]{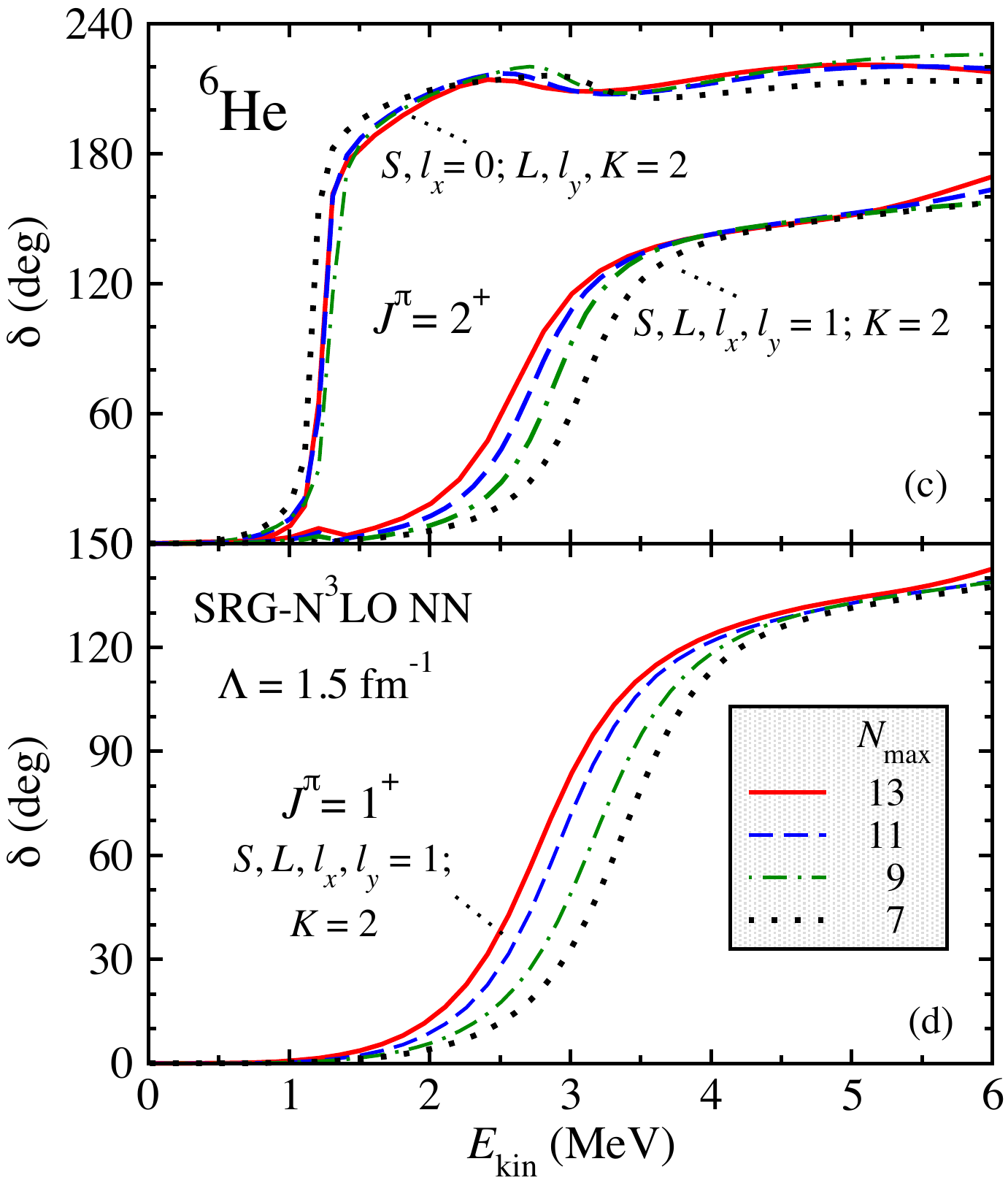}
    \end{minipage}    
\caption{(Color online) Convergence behavior of calculated $^4{\rm He}$+$n$+$n$ (a) $J^\pi = 1^-$ and (b) $0^+$ eigenphase shifts at $K_{\rm max}=19$ and $28$, respectively, and (c) $2^+$ and (d) $1^+$ diagonal phase shifts at $K_{\rm max}=20$ with respect to the size $N_{\rm max}$ of the NCSM/RGM model space. For these calculations we used an extended HO model space of $N_{\rm ext}=70$ and a matching radius of $a=30$ fm. 
In panel (a),  $N_{\rm max}=11$  $1^-$ phase shifts obtained with the $\Lambda=1.8$ fm$^{-1}$ SRG $NN$ potential 
are shown for comparison.}
\label{fig:Nmax}
\end{figure*}
We further describe the $^4$He cluster only by its $I_1^{\pi_1}T_1=0^+0$ g.s.\ and ignore  core polarization effects, which have been estimated to account for $\sim 5\%$ of the $^6$He binding energy~\cite{PhysRevC.88.034320}.  The inclusion of excited states of the core leads to a (presently) unbearable increment of the computational size of the problem. This will be  overcome in the future by  coupling the present  three-cluster model space with  eigenstates of the six-body system within the no-core shell model with continuum (NCSMC)~\cite{PhysRevLett.110.022505,PhysRevC.87.034326}.

We solve Eq.~(\ref{RGMrho}) for the $J^{\pi}=0^{\pm},1^{\pm}$  $2^{\pm} $ channels, and extract the corresponding phase shifts from the diagonal elements of the three-body scattering matrix or from its diagonalization, when large off-diagonal couplings are present.
A summary of the obtained low-lying attractive phase shifts is presented in Fig.~\ref{fig:final}. We have identified several resonances.
The lowest and sharpest appears in the $2^+$ channel around 1.25 MeV
above the $^4$He+$n$+$n$ threshold. An analysis of this resonance, corresponding to the very well known first excited state of $^6$He, shows that it is dominated by $^1S_0$ neutrons in an $\ell_y=2$ relative motion with respect to the $^4$He g.s.\ ($S,\ell_x=0;\;L,\ell_y,K=2$).
A second broader $2^+$ resonance emerges at $\sim2.6$ MeV, where the prevalent picture is that of the halo neutrons with aligned spins, moving relative to
each other and to the core in $P$ wave ($S,\ell_x,L,\ell_y=1;\;K=2$). The same structure also characterizes a $1^+$ resonance located at slightly higher energy.
Resonances also appear in the $2^-$ and $0^-$ channels, dominated by $S,\ell_x,L=1$ and $\ell_y=0$ quantum numbers.
On the other hand, the homogeneous growth through 90$^\circ$ characteristic of a resonance is not present in the 1$^-$ or in the $0^+$ channels. Therefore we cannot see any evidence of a low-lying state that could
be identified with the $1^-$ soft dipole mode suggested in Refs.~\cite{PhysRevLett.85.262} and~\cite{Nakamura2000209}. In addition, our results do not support the presence of a low-lying $0^+$ monopole resonance above the $1^+$ state reported by previous theoretical investigations of the $^4$He+$n$+$n$ continuum, in which the $^4$He was considered as an inert particle with no 
structure. These three-body calculations, performed within the hyperspherical harmonics basis~\cite{PhysRevC.55.R577,Danilin1998383,Ershov:1997az,Descouvemont:2005rc} and with the complex scaling method~\cite{Aoyama:1995,Aoyama:1995b},  obtained a similar sequence of $2^+_1$, $2^+_2$, $1^+$ and $0^+_2$ levels, but different resonance positions and widths. (Only the first two $2^+$ resonances were shown in Ref.~\cite{Descouvemont:2005rc}.) Microscopic $^4$He+$n$+$n$ calculations based on schematic interactions were later reported in Refs.~\cite{Korennov2004249, PhysRevC.80.044310}, but showed only results for the $2^+_1$ narrow resonance and do not comment on a $0^+$ excited state. 

In Fig. \ref{fig:spec}, the energy spectrum of states extracted from the resonances of Fig.~\ref{fig:final} is compared
to the one recently measured at GANIL~\cite{Mougeot:2012aq}. Our results are consistent with the presence of the second low-lying narrow $2^+$ resonance observed for the first time in this experiment. 
A $J=1$ resonance was also measured at 4.3 MeV, however, the parity of such state is not yet determined
and is not possible to univocally identify it with the $1^+$ resonance found at 
2.77 MeV in the present calculations. At the same time, the energy-dependence of the $1^-$ eigenphase shifts of Fig.~\ref{fig:final}(b) does not favor the interpretation of this low-lying state as a dipole mode. We also predict two broader negative-parity states not observed.

A thorough study of the convergence of the results with respect to all parameters defining the size of our model space was performed. These are the maximum value $K_{\rm max}$ of the hyperangular momentum in the expansion~(\ref{hyper_expansion}); the size $N_{\rm max}$ of the HO basis used to calculate the g.s.\ of $^4$He and the localized parts of Eqs.~(\ref{eq:Hkernel}) and~(\ref{eq:Nkernel}); 
and finally, the size $N_{\rm ext}\gg N_{\rm max}$ of the extended HO basis used to represent a 
delta function in the core-halo distance entering the portion of the Hamiltonian kernel that accounts for the interaction between the halo neutrons (see Eq.~(39) of Ref.~\cite{PhysRevC.88.034320}).
In each case the number of integration points and the hyperradius $a$ used to match internal and asymptotic solutions within the $R$-matrix method on Lagrange mesh were chosen large enough to reach stable, $a$-independent results. 
All calculations were performed with the same $\hbar\Omega=14$ MeV frequency adopted for the study of the $^6$He g.s.~\cite{PhysRevC.88.034320}.

We first set the extended HO basis size to the value ($N_{\rm ext}=70$) we found to be sufficient for the $0^+$ g.s.\ energy~\cite{PhysRevC.88.034320}, 
and established that expansion~(\ref{hyper_expansion})
converges at $K_{\rm max}=19/20$ for all negative/positive-parity channels except the $0^+$, requiring $K_{\rm max}=28$. Examples of the convergence pattern with respect to the HO basis size $N_{\rm max}$ are shown in 
Fig.~\ref{fig:Nmax}. In general convergence is satisfactory at $N_{\rm max}$=13. For the higher-lying resonances this value 
is not quite sufficient, but already provides the qualitative behavior to start discussing the continuum structure of the system.    
\begin{figure}[t]
\includegraphics[width=75mm]{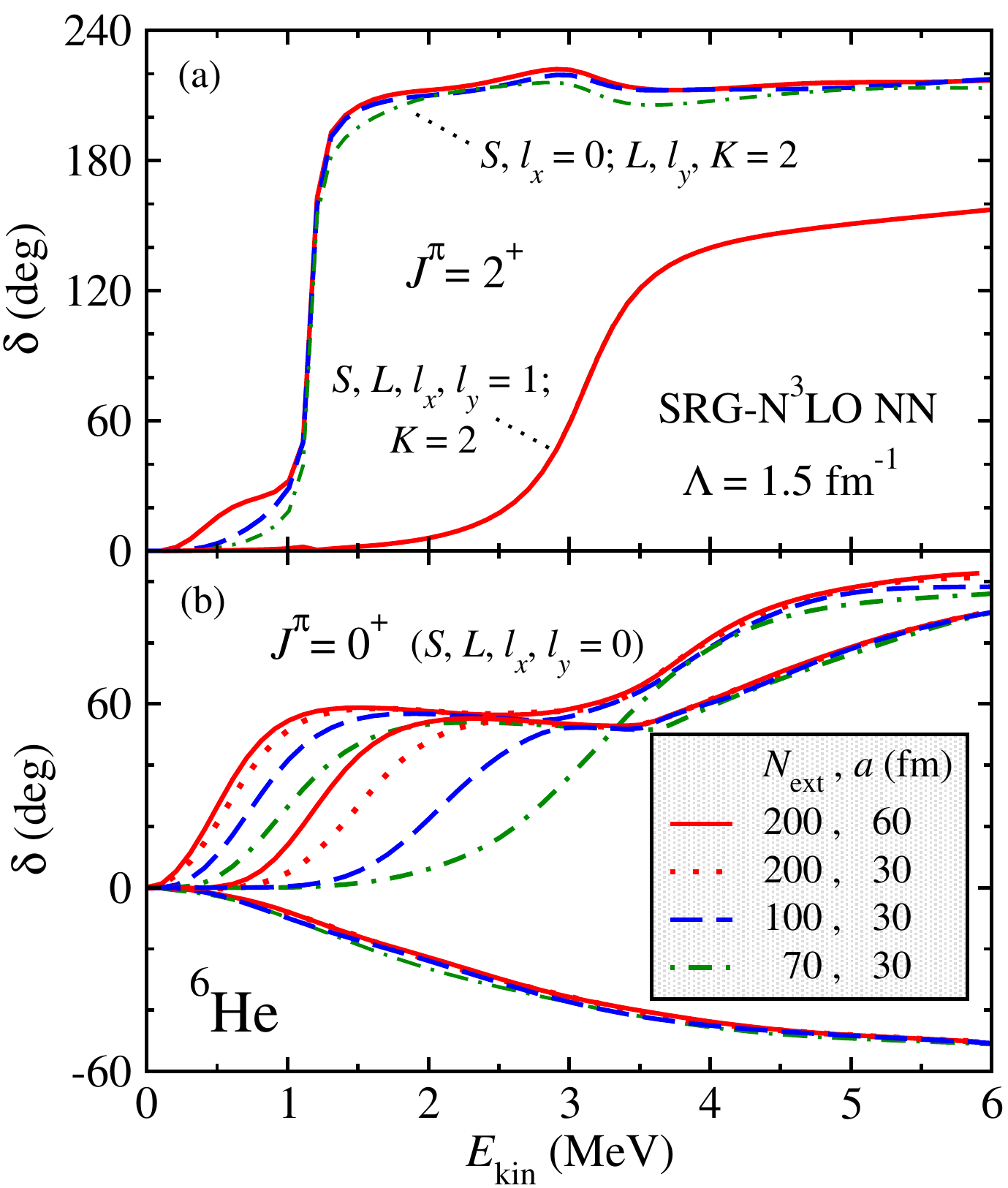}
\caption{(Color online) Dependence on the size of the extended HO model space $N_{\rm ext}$ 
of calculated $^4{\rm He}$+$n$+$n$ (a) $J^\pi = 2^+$ diagonal phase shifts at $N_{\rm max}=7$, and (b) $0^+$ eigenphase shifts at $N_{\rm max}=13$. The curves overlap for the $2_2^+$ resonance.} \label{fig:Next}
\end{figure}
Next we study the dependence on $N_{\rm ext}$, which regulates the range of the potential kernel. 
Not unexpectedly, an increase of $N_{\rm ext}$ requires
at the same time incrementing the matching hyperradius $a$ needed to reach the
asymptotic region (we used values of up to 60 fm) and $K_{\rm max}$, for which we used values as high as 40 in the $0^+$ channel. This limited the maximum value of $N_{\rm ext}$ used to obtain our best ($N_{\rm max}=13$) results for the $0^+,1^-$, and $2^+$ results of Figs.~\ref{fig:final} and~\ref{fig:spec} to 200, $110$, and $90$, respectively. 
As shown in Fig.~\ref{fig:Next}, the influence of $N_{\rm ext}$ is most pronounced for attractive phase shifts in which the two neutrons are in $^1S_0$ relative motion. There the $nn$ interaction is larger and the wave function is more extended due to the Pauli exclusion principle. By far the dominating effect is the steeper onset of the $0^+$ attractive eigenphase shifts that, as already noted in Ref.~\cite{Descouvemont:2005rc}, becomes more accentuated for (higher-lying) components with $K>0$ . However, the qualitative results remain unchanged. In particular, the value of $N_{\rm ext}$ has little or no influence on the position and width of the resonances.
Also, the binding energy of the $0^+$ ground
state of $^6$He calculated in \cite{PhysRevC.88.034320} remains 
unchanged within this much larger model space. Finally, 
changing the value of the SRG parameter used to soften the $NN$ interaction to $\Lambda=1.8$ fm$^{-1}$ does not change the overall structure of the continuum states. Bearing in mind that with this harder
potential convergence is slower, 
in each channel 
we obtain the same number of resonances with similar widths, though somewhat shifted in energy (less than 1 MeV), as shown in Fig.~\ref{fig:Nmax} for the $1^-$. 
This is evidence that the softness of the potential used is not introducing any spurious resonances
and, therefore, verifies the reliability of our results.
\paragraph{Conclusions.}
We calculated, for the first time within an \emph{ab initio} 
approach, the continuum spectrum of
$^6$He  as a $^4$He+$n$+$n$ system. Given the low two-neutron separation energy of this nucleus, including the three-cluster basis in the
calculation is essential. 
We found several resonances, including the well-known narrow $2_1^+$ and the recently measured broader
$2_2^+$. Additional resonant states emerged in the $2^-$, and $1^+$ channels near the second $2^+$ resonance, and in the $0^-$ at slightly higher energy. We found no evidence of low-lying resonances in the $0^+$ and $1^-$ channels.
Therefore, our results do not support the idea that the accumulation of dipole strength at low energy is
originated by a three-body $1^-$
 resonance.

The inclusion of $3N$ forces and core polarization effects  through the NCSMC coupling are underway and will increase 
the predictive capability of the method.
Finally, we expect that complementing this approach with the use of
two integral relations derived from the Kohn variational principle~\cite{Barletta:2009es,RomeroRedondo:2011ci,kievs2010} will increase 
the range of systems that can be described by limiting the distance for which the wave function has to be calculated. 
This will be
essential for the study of $^{11}$Li within a $^9$Li+$n$+$n$ basis.
 
\begin{acknowledgments}
Computing support for this work came from the LLNL institutional Computing Grand Challenge
program and from an INCITE Award on the Titan supercomputer of the Oak Ridge Leadership Computing Facility (OLCF) at ORNL. We thank the Institute for Nuclear Theory at the University of Washington for its hospitality and the Department of Energy for partial support during the completion of this work. Prepared in part by LLNL under Contract DE-AC52-07NA27344. Support from the 
NSERC Grant No. 401945-2011 and U.S. DOE/SC/NP (Work Proposal No. SCW1158)  
is acknowledged. TRIUMF receives funding via a contribution through the 
Canadian National Research Council. 
\end{acknowledgments}

\bibliographystyle{apsrev4-1}

%

\end{document}